\documentclass[aps,prb,twocolumn,showpacs,floatfix]{revtex4-1}
\usepackage{graphicx}
\usepackage{dcolumn}
\usepackage{amsmath,amssymb,amsthm}
\usepackage{color}

\begin{document}

\title{Light-hole exciton mixing and dynamics in Mn-doped quantum dots}
\author{V. Moldoveanu$^{1}$, I. V. Dinu$^1$, R. Dragomir$^{1,2}$ and B. Tanatar$^3$}
\address{$^1$ National Institute of Materials Physics, PO Box MG-7, Bucharest-Magurele,Romania}
\address{$^2$ Faculty of Physics, University of Bucharest, Romania}
\address{$^3$ Department of Physics, Bilkent University, Bilkent, 06800 Ankara, Turkey}
\date{\today}

\begin{abstract}

We investigate theoretically the spectral and dynamical effects of the short-range exchange interaction between a single
manganese (Mn) atom hosted by cylindrical CdTe quantum dots and its light-hole excitons or biexcitons. Our approach is
based on the Kohn-Luttinger $k\cdot p$ theory and configuration interaction method, the dynamics of the system in the
presence of intraband relaxation being derived from the von Neumann-Lindblad equation. The complex structure of the
light-hole exciton absorption spectrum reveals the exchange-induced exciton mixing and depends strongly on the Mn position.
In particular, if the Mn atom is closer to the edges of the cylinder the bright and dark light-hole excitons are mixed
by the hole-Mn exchange alone. Consequently, their populations exhibit exchange-induced Rabi oscillations which can
be viewed as optical signatures of light-hole spin reversal. Similar results are obtained for mixed biexcitons,
in this case the exchange-induced Rabi oscillations being damped by the intraband hole relaxation processes. 
The effect of light-hole heavy-hole mixing is also discussed.

\end{abstract}


\maketitle

\section{Introduction}

More than a decade ago photoluminescence spectroscopy measurements \cite{BesombesPRL2004} revealed
the six excitonic emission lines of CdTe quantum dots (QD) doped with manganese (Mn) ions. These lines were
unambiguously associated with $2M+1$ projections on the $z$-axis of the  Mn spin $M=5/2$ and provided a direct
evidence of its exchange interaction with electrons and holes. Nowadays, the optically active QDs
with a single magnetic dopant are intensively studied as promising solotronic devices 
\cite{KoenraadNatureMaterials2011,KobakNatureCommunications2014}.

A series of recent experiments support these ambitious expectations. The coherent precession of a localized Mn
spin embedded in a CdTe quantum dot has been recently probed \cite{Goryca} and the exciton luminescence of CdTe/ZnTe QDs
 in the presence of a cobalt ion was observed \cite{KobakNatureCommunications2014}. The formation of singlet 
and triplet states in self-assembled heavy-hole (HH) QDs in the presence of a Mn atom has been discussed recently 
\cite{Trojnar,Mendes}. The experimental studies of L\'{e}ger {\it et al.} \cite{Leger}
revealed that the in-plane QD asymmetry and Mn position competitively contribute to the level structure of the Mn `dressed'
excitons and lead to very different photoluminescence spectra. Spin population trapping and optically dressed states
under strong resonant excitation have been reported \cite{Jamet}.
A study on the valence band mixing of heavy hole excitons has been performed by Kyrychenko 
and Kossut \cite{Kyrychenko}.

Bree {\it et al.} \cite{Bree} emphasized the different optical properties of Mn-doped III-V QDs (e.g InAs) due
to the acceptor character of the magnetic atom \cite{Kudelski}. Later on Thuberg {\it et al.} \cite{Thuberg}
investigated theoretically the ultrafast light-induced dynamics in  Mn-doped InAs QD.

The enhanced manipulation of the Mn spin via optical
Stark effect was addressed both experimentally \cite{LeGallPRL2011} and theoretically \cite{ReiterPRB2012}.
Varghese {\it et al.} \cite{Varghese} prepared a positively charged exciton and then recorded the dynamics of the Mn spin
when interacting with a single hole. Quite recently, Pakuski  {\it et al.} \cite{Pakuski} tuned the exciton emission spectrum
of a single Mn-doped QD embedded in a micropillar cavity to its single mode. This opens the way to photonic devices based
on magnetic impurities.

The non-trivial effects of the Mn-exciton exchange interaction are rooted in the simultaneous spin-flip processes
of the Mn and electron or hole spins. Typically, the fully interacting exciton states of the magnetically doped system
become mixtures of `exchange-free' bright and dark excitons (i.e exciton states calculated in the absence of exchange interaction;
for details and notation see Section II). When calculated theoretically, the time-dependent populations of the latter states
develop the so-called exchange-induced Rabi (EXR) oscillations. Reiter {\it et al.} \cite{ReiterPRL2009} exploited this
mechanism and simulated the optical switching of the Mn spin by optically addressing a heavy-hole exciton.

For purely HH systems the electron-Mn (e-Mn) exchange is fully responsible for the appearance of EXR oscillations. It couples bright
and dark excitons with opposite electron spins \cite{Manaselyan} while the hole-Mn (h-Mn) interaction contributes only to the splitting of the six absorption peaks via the diagonal Ising part. The heavy-hole spin flip processes are prevented by selection rules
\cite{selection}.
In turn, a light-hole (LH) exciton experiences both electron and hole spin flips \cite{Reiter-LH}. In this case the exciton states of
the QD-Mn system are mixtures of more than two `exchange-free' excitons, which at first glance prevents one to
distinguish effects of the h-Mn exchange alone or to selectively flip the LH spin whereas conserving the electronic spin.
The optical generation of LHs by highly focused optical-vortex beams has been theoretically investigated in a recent work
\cite{Quinteiro}.

In this work we consider from a theoretical point of view two issues which to our best knowledge were not systematically
studied for Mn-doped QDs supporting light-hole excitons: (i) the role of the Mn position on the exciton or biexciton mixing
as well as on absorption spectra and (ii) the h-Mn  exchange-induced Rabi oscillations of LH excitons and biexcitons.

Let us note that previous studies on electronic structure \cite{Qu,Nguyen} or purely HH systems \cite{Govorov} emphasized
that the exchange interaction with the Mn atom depends strongly on its position and changes the energy spectrum.
Our calculations illustrate how the position dependent e-Mn and h-Mn exchange controls the light-hole exciton mixing which
in turn changes the absorption spectra and the EXR oscillations. Let us stress that the Mn location can be experimentally
controlled \cite{Gietka}.


It is well known that a light-hole ground state in optically active QDs can be experimentally achieved either by tuning
the aspect ratio (i.e height/radius ratio) \cite{Climente,Zielinski} or by applying tensile strain on the structure \cite{Huo}.
Here we take the first route and investigate QDs whose dominant light-hole character of ground state excitons is controlled
by the aspect ratio.

The rest of the paper is organized as follows: Section II presents the essential features of our model and formalism,
Section III contains the results on LH excitons (subsection A) and mixed biexcitons (subsection B). The conclusions are
given in Section IV.

\section{The model}

The starting point of our study is the calculation of the single-particle wavefunctions for electrons and holes
confined in a cylindrical QD. To this end we use the four-band Kohn-Luttinger (KL) Hamiltonian for valence band states
and the $k\cdot p$ theory (see Ref.~\onlinecite{PRB-X} for more details). The confinement potential
leads to the following definition of the envelope functions ($\rho,\theta$ and $z$ are cylindrical coordinates):
\begin{equation}
\phi_{m_znl}(\rho,\theta,z)=\frac{e^{im_z\theta}}{\sqrt \pi R}\,\frac{J_{m_z}(\alpha_n^{m_z}\rho/R)}
{|J_{m_z+1}(\alpha_n^{m_z})|}\,\xi_l(z).\label{Bessel}
\end{equation}
Here $m_z$ is the orbital quantum number, $\alpha_n^{m_z}$ is the $n$-th zero of the Bessel function $J_{m_z}$ and $\xi_l$
are eigenfunctions associated with the vertical confinement, that is $\xi_l(z)=\sqrt{\frac{2}{H}}\sin\left(\frac{\pi lz}{H}\right )$
for $l$ even and $\xi_l(z)=\sqrt{\frac{2}{H}}\cos\left(\frac{\pi lz}{H}\right )$ for $l$ odd.
The QD radius and height are denoted by $R$ and $H$, respectively. The hole states are linear combinations of
basis vectors $|J_z,m_z,n,l\rangle:=|\phi_{m_znl}\rangle |J_z \rangle$:
\begin{equation}\label{spinors}
|\psi^h_{i}\rangle = \sum_{J_z+m_z=F_z}\sum_{n,l}C^{i,F_z}_{n,l}|\phi_{m_znl}\rangle |J_z \rangle,
\end{equation}
where we introduced the total orbital quantum number $F_z=J_z+m_z$ and the Bloch band-edge states $|J_z\rangle$.
Heavy (light) hole states correspond to $J_z=\pm{3/2}$ ($J_z=\pm{1/2}$). The electrons in the conduction band are
described by a single-band effective mass Hamiltonian, its eigenstates $|\psi^e_j\rangle$ being described
by $|\phi_{m_z^jn^jl^j}\rangle |S_z^j \rangle$ where $S_z=\pm1/2$ is the electron spin.
The energies in the conduction band are denoted by $E^c_i$ and the hole energies by $E_j^h$.

At the next step we use the single-particle functions to calculate many-body interacting configurations (excitons, biexcitons etc.)
and energies in the absence of the Mn atom. The QD is described by the Hamiltonian $\hat{H}_0$:
\begin{eqnarray}\nonumber
&&\hat{H}_0=\sum_{i=1}^{N_C}E_{i}^{c}a_{i}^{\dagger}a_{i}+\sum_{j=1}^{N_V}E_{j}^{h}b_{j}^{\dagger}b_{j}
+\frac{1}{2}\sum_{i,j,k,l} a_{i}^{\dagger}a_{j}^{\dagger}a_{l}a_{k}V_{ijkl}^{ee} \\\nonumber
&&+\frac{1}{2}\sum_{i,j,k,l} b_{i}^{\dagger}b_{j}^{\dagger}b_{l}b_{k}V_{lkji}^{hh}
-\sum_{i,j,k,l}a_{i}^{\dagger}b_{l}^{\dagger}b_{j}a_{k}(V_{ijkl}^{eh}-V_{ijlk}^{eh}),\\\label{HHzero}
\end{eqnarray}
written in terms of creation/annihilation operators for electrons and holes. The single-particle indices $i$ and $j$ are
restricted to $N_C$ and $N_V$ in view of numerical diagonalization via configuration interaction method 
(more details are given in Section III).

In Eq.(\ref{HHzero}) we singled out the direct ($V_{ijkl}^{eh}$) and exchange ($V_{ijlk}^{eh}$) electron-hole interactions. 
The latter has both short-range and long-range components. The splitting between the two pairs of bright and dark excitons
is mainly given by the short-range exchange \cite{Kadantsev} while the anisotropic (non-local) long-range exchange
contributes to the fine structure splitting (FSS) and mixing of the bright excitons. Under strong tensile strain the
anisotropic exchange interaction increases and one can switch the ground state of a QD from heavy-hole to light-hole \cite{Huo}.
For isotropic QD confinement potential the bright exciton splitting vanishes \cite{Kadantsev,Lin}. 
Moreover, for the QDs considered here the strain is assumed to be small and therefore the FSS is neglected. 
The calculated eigenstates and eigenvalues of $\hat{H}_0$ are denoted by $|\nu\rangle$ and  ${\cal E}_{\nu}$.

A useful description of the many-body states (MBS) $|\nu\rangle$ is given by the spins of the electrons and holes
which occupy well defined single-particle states. For example, the LH exciton state containing a spin-up electron
in the conduction band and a light-hole of spin $J_z=-1/2$ is denoted by $|\uparrow\Downarrow_L\rangle$. Similarly,
$|\downarrow\Uparrow_H\rangle$ stands for a HH exciton of spin $J_z=3/2$. We stress that if HH-LH mixing is present
$J_z$ denotes the dominant component in the Luttinger spinors given by Eq.(\ref{spinors}).

Recent studies emphasized that a small built-in strain induces magnetic anisotropy \cite{Varghese} and a fine structure of the
Mn levels. Formally this is described by the term $D_0\hat{M}_z^2$ and one has:
\begin{equation}\label{Hzero}
(\hat{H}_0+D_0\hat{M}_z^2)|\nu,M_z\rangle=({\cal E}_{\nu}+D_0M_z^2)|\nu,M_z\rangle,
\end{equation}
where $M_z$ is the $z$ projection of the manganese spin ($M_z=\pm\frac{5}{2},\pm\frac{3}{2},\pm\frac{1}{2}$)
and ${\hat M}_z|M_z\rangle=M_z|M_z\rangle$.

The short-range exchange interaction between the electron (hole) spins $\vec{S}$ ($\vec{J}$) and the manganese spin
$\vec{M}$ located at $\bf{R_{{\rm Mn}}}$ is given by:
\begin{eqnarray}\nonumber
H_{{\rm X-Mn}}&=&-J_e\vec{S}\vec{M}\delta({\bf r}_e-{\bf R}_{{\rm Mn}})+J_h\vec{J}\vec{M}\delta({\bf r}_h-{\bf R}_{{\rm Mn}})\\\label{H_Mn1}
&=&H_{{\rm e-Mn}}+H_{{\rm h-Mn}}.
\end{eqnarray}
where $J_e$ and $J_h$ are the e-Mn and h-Mn exchange interaction strengths.

It is convenient to write down the second quantized form $\hat{H}_{{\rm X-Mn}}$ of $H_{{\rm X-Mn}}$ in the basis $\{|\nu,M_z\rangle\}$.
The matrix elements of e-Mn and h-Mn exchange are found by standard calculation. For example:
\begin{eqnarray}\nonumber
&&\langle\nu,M_z|\hat{H}_{{\rm e-Mn}}|\nu',M_z'\rangle=
-J_e\sum_{i,j=1}^{N_C}\overline{\psi^e_i}({\bf R}_{{\rm Mn}})\psi^{e}_j({\bf R}_{{\rm Mn}})\times\\\label{e-Mn-2Q}
&&\langle S_z^iM_z|\vec{\hat{S}}\vec{\hat{M}}|S_z^jM_z'\rangle
\langle\nu|a_i^{\dagger}a_j|\nu'\rangle .
\end{eqnarray}
The product $\vec{S}\vec{M}$ in Eq.\,(\ref{e-Mn-2Q}) contains spin flip operators $\hat{S}_{\mp}\hat{M}_{\pm}$
which mix the `exchange-free' states $|\nu,M_z\rangle$, and the Ising part $\hat{S}_z\hat{M}_z$ which would simply turn each state
$|\nu\rangle$ into a manifold $|\nu,M_z\rangle$ with a fine
structure controlled by $J_e$. Note that the matrix elements of the exchange interaction depend explicitly on the
electronic wavefunctions evaluated at the Mn position. The hole-Mn exchange acquires a more complicated expression
because a given Luttinger spinor is generally a combination of LH and HH states. Consequently for a pair of states
$\{J_z,J_z' \}$ the matrix elements $\langle J_zM_z|\vec{\hat {J}}\vec{\hat {M}}|J_z'M_z'\rangle$ must be multiplied by
their weights in the Luttinger spinor.

The total Hamiltonian of the QD-Mn system finally reads as:
\begin{equation}\label{HTotal}
\hat{H}=\hat{H}_0+D_0\hat{M}_z^2+\hat{H}_{{\rm e-Mn}}+\hat{H}_{{\rm h-Mn}}.
\end{equation}

 The eigenstates of $\hat{H}$  can be written as:
\begin{equation}\label{Mn-states}
|p\rangle=\sum_{\nu,M_z}A^p_{\nu,M_z}|\nu,M_z\rangle,
\end{equation}
where the coefficients $A^p_{\nu,M_z}$ are found by diagonalization. For further use we introduce the
 weight of the `exchange-free' state $|\nu,M_z\rangle$ in a fully interacting configuration 
$|p\rangle$ as $w^p_{\nu,M_z}:=|A^p_{\nu,M_z}|^2$.

The light-matter interaction is treated classically, the corresponding Hamiltonian being:
\begin{equation}
\hat{V}_R(t)=\frac{eA_0(t)}{m_0}\sum_{i,j}\left ((e^{-i\omega t}S^+_{ij}+e^{i\omega t}S^-_{ij})a^*_ib_j+h.c \right ),
\end{equation}
where $m_0$ is the free electron mass, $A_0(t)$ is the pulse envelope and $\omega$ is its frequency. 
We shall denote by $E$ the electric field corresponding to the vector potential $A_0(t)$.
For simplicity we considered real rectangular envelopes. $S^{\pm}_{ij}$ denotes the interband optical coupling
matrix elements ($\vec{e}_{\sigma_{\pm}}$ are right/left circular polarization vectors):
\begin{equation}\label{SS}
S^{\pm}_{ij}=\langle \psi_i^e,(\vec{p}\vec{e}_{\sigma_{\pm}})\psi_j^h \rangle.
\end{equation}
Note that in the basis $\{\nu,M_z\}$ the operator $\hat{V}_R(t)$ is diagonal with respect to $M_z$ and off-diagonal with respect to
$\nu$. $S^{\pm}_{ij}$ are calculated in terms of the parameter $P=\frac{i\hbar}{m_0}\langle s,p_x X\rangle$ using known
values for the Kane energy\cite{Gywat} $E_p=\frac{2}{m_0}|\langle s,p_x X\rangle|^2=\frac{2m_0}{\hbar^2}P^2$.

The dynamics of the many-body configurations under ultrashort laser pulses is derived from the density operator
$\hat{\rho}$ of the system which obeys the von Neumann-Lindblad equation:
\begin{equation}\label{vNL}
i\hbar {\dot{\hat\rho}}(t)=[\hat{H}+\hat{V}_R(t),\hat{\rho}(t)]+i\sum_{\lambda }{\cal L}_{\lambda}[\rho(t)],
\end{equation}
where the dissipative operator ${\cal L}_{\lambda}$ describes the intraband relaxation for electrons ($\lambda=e$) and holes
($\lambda=h$) and is given by ($\{,\}$ denotes the anticommutator):
\begin{eqnarray}\nonumber
{\cal L}_{\lambda}[\rho(t)]=\sum_{i,j=1}^{N_{\lambda}}\gamma_{ij}
\left\lbrace X_{ij}^{\lambda}\rho(t) X_{ij}^{\lambda\dagger}-\frac{1}{2}
\{X_{ij}^{\lambda\dagger}X_{ij}^{\lambda},\rho(t)\}\right\rbrace.
\end{eqnarray}
The relaxation rates $\gamma_{ij}$ are associated to a pair of single-particle states in the band $\lambda$ and the
jump operators are defined as $X_{ij}^{e}=a^{\dagger}_ia_j$ and $X_{ij}^{h}=b^{\dagger}_ib_j$.
The spin-conserving hole relaxation is the fastest process described by the relaxation time for holes
$\tau_h=\gamma_{ij}^{-1}$ where $i,j$ are hole levels having the same spin. Spin-conserving relaxation for
electrons in the conduction band will not be considered here as the numerical simulations presented in Section
II involve only the lowest single-particle states from the conduction band. The hole spin relaxation is neglected
as it is a much slower process.

The population of a given state $|\nu,M_z\rangle$ is given by the diagonal matrix element
$P_{\nu,M_z}:=\langle\nu ,M_z|\hat{\rho}(t)|\nu,M_z\rangle$ whereas the coherence between two states is
given by off-diagonal elements. The statistical average of the $z$-projection of the Mn spin is given by:
\begin{equation}
\langle \hat{M_z} \rangle=\sum_{\nu,M_z} M_z\langle\nu ,M_z|\hat{\rho}(t)|\nu,M_z\rangle .
\label{medieMz}
\end{equation}

For a fixed initial configuration, the von Neumann-Lindblad equation is numerically solved
on a truncated Fock subspace containing the states involved in optical transitions or spin-flip processes.

\section{Results and discussion}

In this work we consider cylindrical QDs whose single-particle hole ground state achieves a dominant LH character by
appropriate tuning of the aspect ratio $H/2R$. For example, we find that for $R=5$\,nm and $H=11$\,nm the
highest energy single-particle valence states of the KL Hamiltonian are up to $86\%$ made of $s$-shell 
light holes ($F_z=J_z=\pm 1/2$) and the next two states are mostly ($82\%$) $s$-shell HH-like ($F_z=\pm 3/2$). 
Due to the valence band mixing the 'mostly' LH Luttinger spinor has smaller $p$-shell HH weights 
as well (e.g. a $10\%$ HH component $m=-1,J_z=3/2$) but the total quantum number $F_z=J_z+m_z$ is conserved.
The existence of a LH ground state for tall cylindrical quantum dots is also confirmed by atomistic tight-binding 
calculations \cite{Zielinski}.

For our system the gap between the higest energy LH and HH doublets is approximatively 7\,meV.
It also turns out that the next hole states which could be optically coupled to the
$s$-shell electronic states are located 27\,meV below the highest energy HH states while the next electron 
states are 80\,meV above the lowest energy $s$-shell doublet. Systematic numerical simulations show that 
by enlarging $N_V$ and $N_C$ the energies of the exciton or mixed biexciton states considered in this work
 are modified  by only a few $\mu$eVs and their mixing with other configurations is negligible.
Typically, a good convergence is already achieved for $N_C=4$ and $N_V=6$. 

The position of the manganese atom is given in cylindrical coordinates by $(\rho_{{\rm Mn}},\theta_{{\rm Mn}},z_{{\rm Mn}})$.
Note that the center of the dot corresponds to $z_{{\rm Mn}}=0$ and therefore $z_{{\rm Mn}}\in [-H/2,H/2]$.
From Eq.\,(\ref{spinors}) one infers that if the Mn atom is at the dot center (i.e $\rho_{{\rm Mn}}=0$)
the $m_z=\pm 1$ states (i.e. $p$-shell states) do not contribute to the exchange interaction as the Bessel functions
$J_{\pm 1}$ vanish at origin. In contrast, by shifting the magnetic impurity from origin the exchange interaction couples $p$-shell
components as well \cite{Qu}. The effect of the valence band mixing will be further discussed in Section III B.

In the numerical calculations we use the following values of the exchange couplings: $J_e=15$\,eV\AA$^3$, $J_h=60$\,eV\AA$^3$.
The Luttinger parameters for CdTe are $\gamma_1=5.37$, $\gamma_2=1.67$, $\gamma_3=1.98$. $D_0$ typically ranges from 0 to few tenths
of $\mu$eV for large strain values \cite{Varghese} and insures a slow relaxation (few $\mu$s) of the Mn spin.
The numerical simulations discussed in this section were obtained for $D_0=6\mu$eV. We found similar results for other
values of $D_0$. The von Neumann-Lindblad equation is solved by taking into account both excitons and biexcitons.

\subsection{Exchange-induced LH exciton mixing and dynamics}

\begin{figure}[tbhp!]
 \centering
 \includegraphics[width=0.49\textwidth]{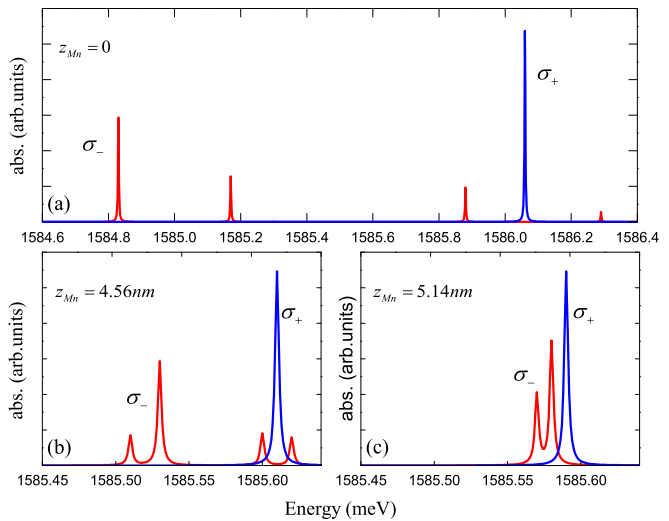}
 \caption{(Color online) Light-hole exciton absorption spectra associated to $\sigma_-$ (red) and $\sigma_+$ (blue)
polarized light for several positions of the Mn atom on the $z$-axis. The initial state of the QD is $|1,5/2\rangle$.}
 \label{Xspectra}
 \end{figure}

 Suppose that the initial state of the QD-Mn system is $|\nu=0,M_z=\frac{5}{2}\rangle$ where $|0\rangle$ is
the QD ground state in the absence of the Mn impurity, that is the valence band states are completely
filled and the conduction band states are empty. The Mn spin can be fixed by applying a magnetic field along
the $z$ axis. Note that one can turn off this magnetic field and still have stable orientation of the Mn spin, due to
rather long relaxation times of the Mn spin.

The `exchange-free' bright LH excitons $|\uparrow\Uparrow_L;5/2\rangle$ and $|\downarrow\Downarrow_L;5/2\rangle$
are generated by circularly polarized $\sigma_+$ and $\sigma_-$ pulses. Then, a peak corresponding to
each exciton appears in the absorption spectrum associated to the `exchange-free' Hamiltonian in Eq.(\ref{Hzero})
(not shown). Using the Fermi Golden Rule one can also calculate the light-hole exciton
absorption spectra of the fully interacting QD-Mn system for several positions of the Mn atom along the $z$-axis,
while keeping $\rho_{{\rm Mn}}=0$.

Fig.~\ref{Xspectra} shows that while the single-peak picture still holds for $\sigma_+$ pulses, a multiple peak structure
develops for $\sigma_-$ polarized light.

More precisely, instead of a single peak
the $\sigma_-$ absorption spectrum displays four peaks for $z_{{\rm Mn}}=0$ (dot center) and  $z_{{\rm Mn}}=4.56$\,nm whereas
for $z_{{\rm Mn}}=5.14$\,nm only two peaks are visible. We also find that if the Mn atom is located even closer to the
cap point $z_{{\rm Mn}}=5.5$\,nm a single $\sigma_-$ peak survives, which roughly coincides with the peak associated to $\sigma_+$
polarization (not shown). Note that the energy range covered by the four $\sigma_-$ peaks shrinks from 1.6\.meV in
Fig.\,\ref{Xspectra}(a) to 0.2\,meV in Fig.\,\ref{Xspectra}(b) while the single $\sigma_+$ peak position does not change significantly as the Mn location
varies.

We find that the complex peak structure is due to the exchange-induced mixing between four non-interacting excitons.
From Eq.\,(\ref{e-Mn-2Q}) one can easily check that $\hat{S}_+\hat{M}_-|\downarrow\Downarrow_L;5/2\rangle=|\uparrow\Downarrow_L;3/2\rangle$ via e-Mn exchange and
$\hat{J}_+\hat{M}_-|\downarrow\Downarrow_L;5/2\rangle=|\downarrow\Uparrow_L;3/2\rangle$ via h-Mn exchange.
Moreover, the latter state pairs with the bright exciton
$|\uparrow\Uparrow_L;1/2\rangle$ due to e-Mn exchange. The mixing of the LH and HH excitons
(e.g. $|\downarrow\Uparrow_L;3/2\rangle$ and $|\downarrow\Uparrow_H;1/2\rangle$) is negligible because of the
few meV gap between these states.

The fully interacting LH excitons can be therefore written as quadruples made of `exchange-free' excitons
and the corresponding absorption peaks depend on the weights of optically active exciton $|\downarrow\Downarrow_L;5/2\rangle$
in a given quadruple. Note however that the $\sigma_+$ exciton cannot be mixed with any other
LH state because $\hat{J}_-\hat{M}_+|\uparrow\Uparrow;5/2\rangle=\hat{S}_{\pm}\hat{M}_{\mp}|\uparrow\Uparrow;5/2\rangle=0$.
Therefore the corresponding absorption peak does not split as the Mn position changes.
The peak position is in turn affected by the non-vanishing Ising part of the exchange interaction (see  Fig.\,\ref{Xspectra}).

\begin{figure}[tbhp!]
 \centering
 \includegraphics[width=0.475\textwidth]{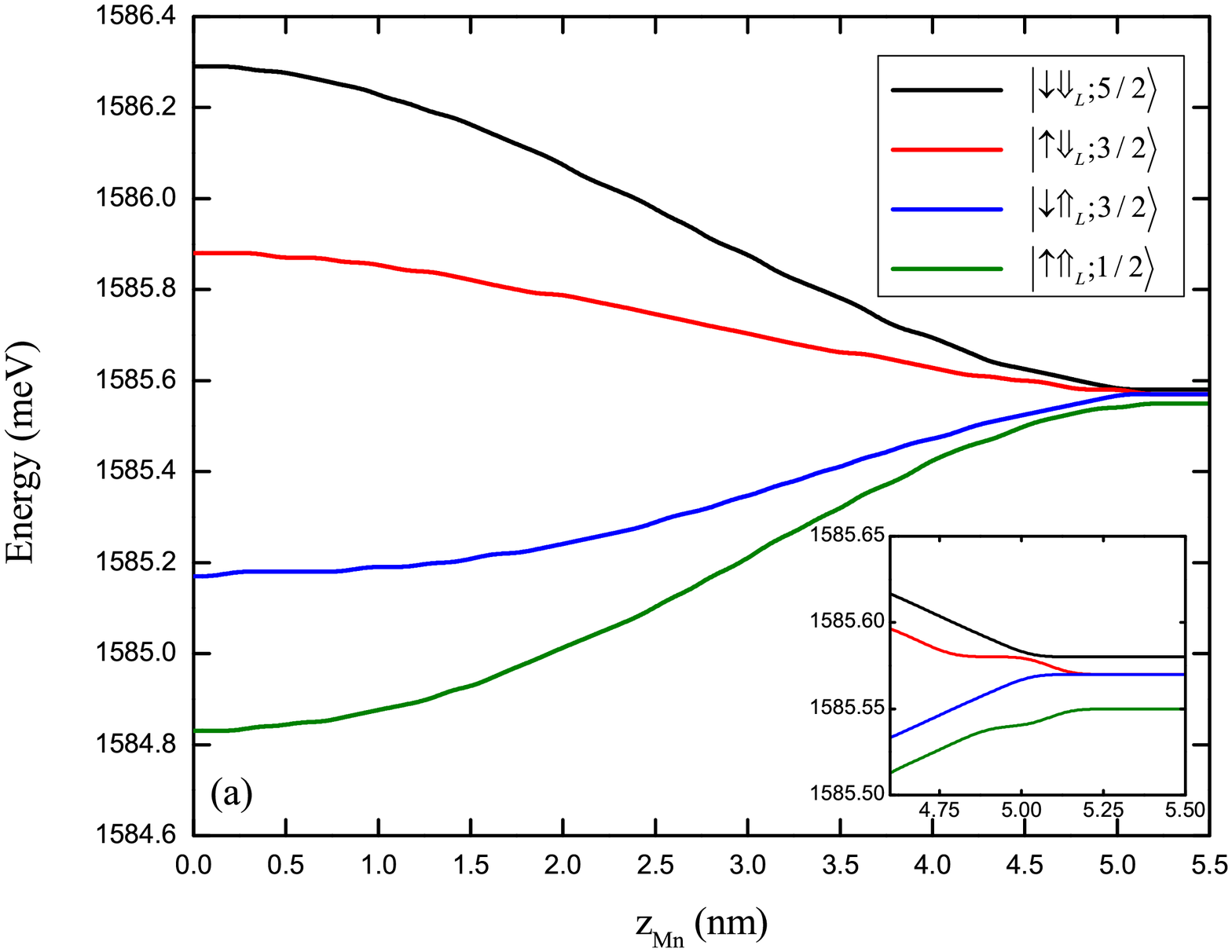}
 \includegraphics[width=0.45\textwidth]{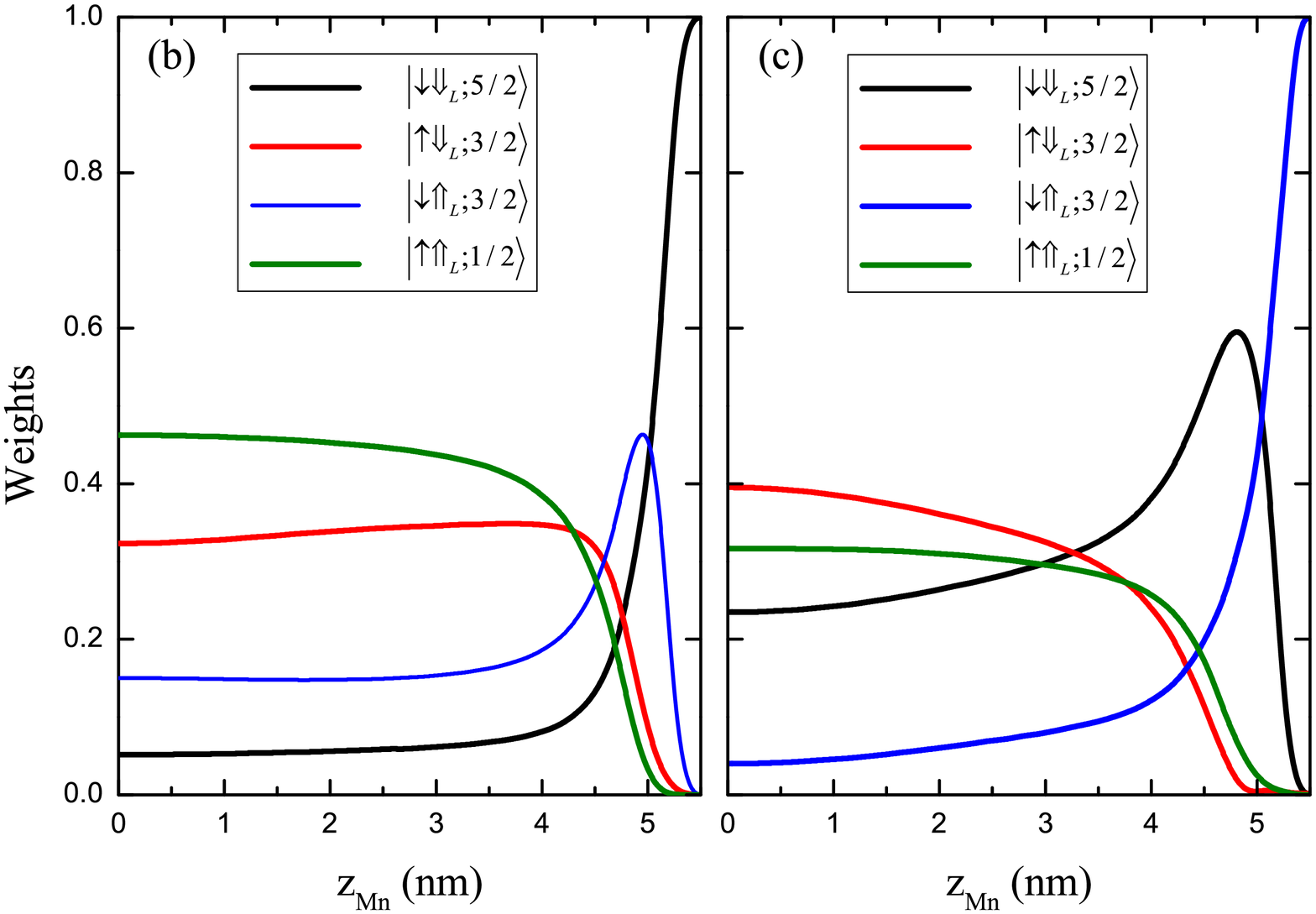}
 \caption{(Color online) (a) The energy levels corresponding to four fully interacting LH excitons as a function of the
Mn position on the $z$ axis. At $z_{{\rm Mn}}=5.5$\,nm each level corresponds to a non-interacting exciton.
(b) and (c) The weights of the `exchange-free' excitons for two fully interacting state as a function of Mn position.
The states evolve from a strongly mixed quadruple to a single bright (b) and dark (c) exciton.  }
\label{Levels}
\end{figure}

In Fig.\,\ref{Levels} (a) we show four LH exciton levels corresponding to quadruples made of the abovementioned `exchange-free'
excitons as a function of the Mn  $z$-coordinate which varies from the cylinder center ($z_{{\rm Mn}}=0$) to its top
edge ($z_{{\rm Mn}}=5.5$\,nm). The electron and hole envelope functions vanish on cylinder surface so the Mn atom and
the excitons no longer interact there.
Then for $z_{{\rm Mn}}=5.5$\,nm the levels in Fig.\,\ref{Levels}(a) correspond to `exchange-free' excitons, and are split only
by the short-range electron-hole exchange and the magnetic anisotropy term $D_0M_z^2$
(see the inset in Fig.\,\ref{Levels} (a)). As $z_{{\rm Mn}}$ approaches the cylinder center the splitting increases and reaches a maximum at $z_{{\rm Mn}}=0$ because the e-Mn and h-Mn exchange interactions attain their largest values at this point.
Indeed, from the explicit expression of the eigenfunction $\xi_l$ below Eq.\,(\ref{Bessel})) one can easily see that
as $z_{{\rm Mn}}$ varies from zero to $\pm H/2$ the off-diagonal matrix elements of the h-Mn interaction
($\langle\downarrow\Uparrow_L;3/2|\hat{H}_{{\rm h-Mn}}|\downarrow\Downarrow_L;5/2\rangle$ and e-Mn interaction
($\langle\uparrow\Downarrow_L;3/2|\hat{H}_{{\rm e-Mn}}|\downarrow\Downarrow_L;5/2\rangle$) decrease and eventually vanish.
Note the clear correspondence between the energy levels and the $\sigma_-$ absorption peaks in Figs.\,\ref{Xspectra}(a) and (b).

To further inspect the effect of the Mn position on the LH mixing excitons and absorption spectrum we
show in Fig.\,\ref{Levels}(b) the weights of the `exchange-free' excitons in the fully interacting state which reduces
to the bright exciton $|\downarrow\Downarrow_L;5/2\rangle$ at $z_{{\rm Mn}}=5.5$\,nm.
The associated level in Fig.~\ref{Levels}(a) is the one with the highest energy and is clearly correlated to the rightmost
$\sigma_-$ peak in the absorption spectrum. A strong and rather constant LH exciton mixing is noticed as long as
$z_{{\rm Mn}}<3.5$\,nm, the dominant states being $|\uparrow\Uparrow_L;1/2\rangle$ and $|\uparrow\Downarrow_L;3/2\rangle$.
The small amplitude of the associated absorption peak (see Fig.\,\ref{Xspectra}(a)) is explained by the negligible
weight of the bright exciton $w_{\downarrow\Downarrow_L;5/2}\sim 0.07$.
From Fig.~\ref{Levels}(b) we notice that at $z_{Mn}=5.14$ nm the weight of the bright exciton increases to 0.60 which
leads to a higher absorption of the rightmost peak in Fig. \,\ref{Xspectra}(c). Moreover, Fig. ~\ref{Levels}(c) 
shows that the bright exciton state has a  substantial weight ($\sim 0.4$) in another fully interacting LH exciton 
state (it actually corresponds to level plotted with blue line in Fig. ~\ref{Levels}(a)). As a consequence the amplitude 
of the second peak from the right in Fig.\,\ref{Xspectra}(c) also increases.

As $z_{{\rm Mn}}$ further approaches the top cylinder edge the fully interacting state turns first to a mixture of
$|\downarrow\Downarrow_L;5/2\rangle$ and $|\downarrow\Uparrow_L;3/2\rangle$ if $z_{{\rm Mn}}\in [4.5:5.25]$\,nm
and finally to the bright exciton $|\downarrow\Downarrow_L;5/2\rangle$. One recognizes at once that the
two states are coupled by the h-Mn exchange only. The weights shown
in Fig.\,\ref{Levels}(c) correspond to the LH state which becomes dark as
$z_{{\rm Mn}}$ approaches the top edge as $w_{\downarrow\Uparrow_L;3/2}$ tends to unity in this limit.

A similar pattern is observed for the remaining two levels in Fig.\,\ref{Levels}(a), that is for $z_{{\rm Mn}}\in (4.5,5.25)$\,nm
the corresponding interacting states are mostly mixtures of $|\uparrow\Downarrow_L;3/2\rangle$ and $|\uparrow\Uparrow_L;1/2\rangle$
(not shown). This suggests that as $z_{{\rm Mn}}$ approaches the cylinder edge the h-Mn interaction dominates
while the e-Mn exchange plays a negligible role.

\begin{figure}[ht]
 \centering
 \includegraphics[width=0.49\textwidth]{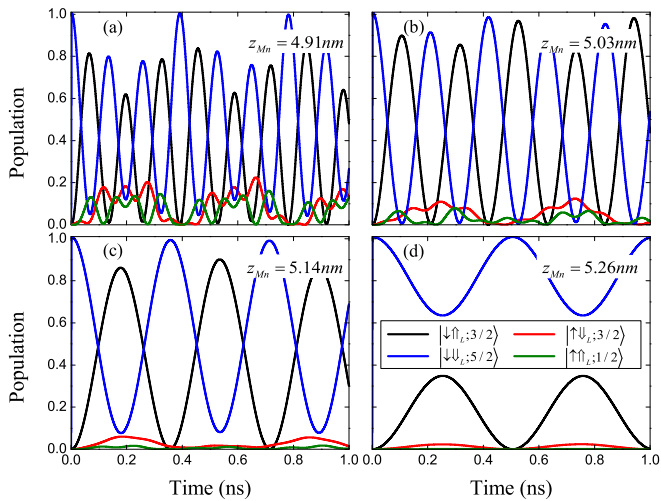}
 \caption{(Color online) Exchange-induced Rabi oscillations of populations of the four `exchange-free' excitons as
a function of the Mn position. The pulse that generates the bright exciton corresponds to an electric field $E=50$\,kV/cm.}
 \label{RabiExciton}
\end{figure}

To explain this behavior one should recall that the e-Mn exchange vanishes much faster than the h-Mn coupling as a function of
$z_{{\rm Mn}}$, since its strength $J_e$ is four times smaller than $J_h$. Moreover, the Ising terms also decrease
if the Mn atom is localized near the cylinder edge. Therefore, by changing the position of the Mn atom one reaches a
regime where the electron spin can hardly be reversed while the hole spin is still subjected to flip-flop processes.
For $z_{{\rm Mn}}<4$nm the electron and hole spin flip processes coexist and one cannot single out the effect of the h-Mn exchange.

As the e-Mn exchange vanishes the quadruple states turn to doublets, the two peaks in Fig.\,\ref{Xspectra}(c)
corresponding to the linear superpositions of $|\downarrow\Downarrow_L;5/2\rangle$ and $|\downarrow\Uparrow_L;3/2\rangle$.

Motivated by previous studies of the exchange-induced Rabi oscillations for purely heavy-hole excitons
\cite{ReiterPRL2009} we also investigated the exchange-induced LH dynamics. We simulate the following two-step setup:
i) the bright $|\downarrow_s\Downarrow_L;5/2\rangle$ exciton is generated via a sharp $\sigma^-$ $\pi$-pulse,
provided the initial state of the system is $|1,5/2\rangle$; ii) then the pulse is turned off and the populations
of the `exchange-free' excitons change only under the e-Mn and h-Mn exchange interaction.

Fig.\,\ref{RabiExciton} shows the populations of the `exchange-free' LH excitons for several positions of the Mn atom.
The EXR oscillations have a much longer period than the $\sigma_-$ pulse and the fast decrease of the ground state
$|1,5/2\rangle$ is not included (not shown). The oscillation amplitude depends only on the exchange interaction strength
and the gap between `exchange-free' excitons. In the strong mixing regime $z_{{\rm Mn}}<4.5$\,nm the exchange-induced Rabi
oscillations display a messy pattern because the e-Mn and h-Mn interactions coexist (not shown) and the interacting
excitons are combinations of four `exchange-free' states. At a formal level this means that in the strong mixing regime
the system dynamics cannot be discussed in the framework of a simple three-level model (i.e. the ground state, bright and dark exciton)
as it is done for HH systems \cite{ReiterPRL2009}. Henceforth, we focus our analysis on the dynamics controlled mostly by the
hole spin flip processes.

Clear exchange induced Rabi oscillations of the states $|\downarrow\Downarrow_L;5/2\rangle$ and $|\downarrow\Uparrow_L;3/2\rangle$
are obtained for $z_{{\rm Mn}}=5.03$\,nm in Fig.\,\ref{RabiExciton}(b), the other two excitons having much smaller populations and
a different oscillation pattern. The oscillation period is quite large (around $0.12$\,ns) as the splitting between the bright and
dark exciton levels is very small (see the inset of Fig.\,\ref{Levels}). The state of the system in this regime is a mixture of the
bright and dark LH excitons. By placing the Mn atom closer to the cylinder edge the mixing of the bright and dark excitons and
the amplitude of the Rabi oscillations decrease as the h-Mn interaction diminishes (see Fig.\,\ref{RabiExciton}(d)).

From the experimental point of view the exchange-induced Rabi oscillations can be indirectly detected using the optical
transition between the bright exciton and the biexciton state $|\uparrow\downarrow\Uparrow_L\Downarrow_L;5/2\rangle$
under a $\sigma^+$ $2\pi$-pulse. The strongest response to such a pulse is expected if it coincides with a maximum of the
bright exciton population.
Then the distance between consecutive higher peaks of the biexciton population roughly equals the period of the Rabi oscillations.

Let us emphasize that the exciton dynamics presented here is not affected by relaxation processes as the optical pulse
addresses only the LH ground state and the lower levels of the valence band are occupied.
Note however that sub-nanosecond recombination processes would eventually dephase the 
exchange-induced Rabi oscillations.

The decoupling of the electron-spin flip and hole-spin flip processes as the Mn
atom approaches the edge is due to the simple fact that the electron-Mn exchange interaction $J_e$ is four times smaller than the
hole-Mn exchange $J_h$. This feature will hold for other confinement potentials (e.g. parabolic).

\subsection{Exchange-induced dynamics of mixed biexcitons}

The biexciton manifold we consider is built from the electronic $s$-shell and the lowest energy LH and HH shells.
As before we start by discussing the non-interacting biexcitons. Apart from purely LH
($|LL\rangle:=|\downarrow\uparrow\Downarrow_L\Uparrow_L\rangle$) and purely HH
$|HH\rangle:=|\downarrow\uparrow\Downarrow_H\Uparrow_H\rangle$
states one finds four `mixed' biexcitons made of both LH and HH states.
 We label these states by the total hole spin as follows ($D$ and $B$ stand for `dark' and `bright'):
$|D+1\rangle :=|\downarrow\uparrow\Downarrow_L\Uparrow_H\rangle$, $|D-1\rangle :=|\downarrow\uparrow\Uparrow_L\Downarrow_H\rangle$,
$|B+2\rangle :=|\downarrow\uparrow\Uparrow_L\Uparrow_H\rangle$ and finally $|B-2\rangle :=|\downarrow\uparrow\Downarrow_L\Downarrow_H\rangle$.

\begin{figure}[ht]
 \centering
 \includegraphics[width=0.475\textwidth]{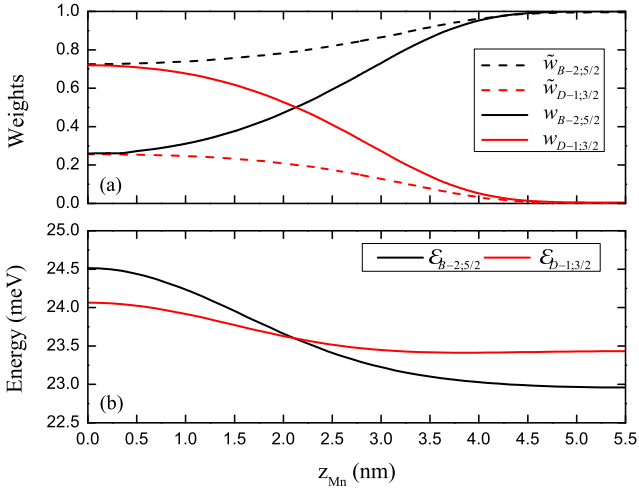}
 \caption{(Color online) (a) The weights of the bright and dark biexcitons as a function of the Mn position. The
solid lines weights are associated to fully interacting states whereas the dashed lines are obtained by neglecting
the Ising term in the Mn-exchange interaction. (b) The bright and dark biexciton levels in the absence of the
flip-spin scattering. The level crossing is only due to the diagonal Ising term of the h-Mn exchange.
Note that the degeneracy point coincides with the point of maximal mixing shown in Fig.\,1(a).}
 \label{B1}
\end{figure}

In the presence of the Mn atom the optically active mixed biexcitons are coupled only by the h-Mn exchange via
light-hole spin flip, as the heavy-hole spin is pinned and the e-Mn is ineffective. We shall focus our discussion on the
two fully interacting states given by the superposition of $|B-2;5/2\rangle$ and $|D-1;3/2\rangle$. The weights of
these states in the fully interacting biexciton as a function of the Mn position are presented in Fig.\,\ref{B1}(a)
(solid lines). We observe that if $z_{{\rm Mn}}$ is closer to the cylinder center the state is mostly dark
(i.e. $w_{D-1,3/2}\sim 0.75$) and that it becomes mostly bright as $z_{{\rm Mn}}$ approaches the cap point.

The strongest overlap of the two states is found at $z_{{\rm Mn}}=2.1$nm although the largest value of the off-diagonal
matrix element $\langle B-2;5/2|\hat{H}_{h-{\rm Mn}}|D-1;3/2\rangle$ corresponds to $z_{{\rm Mn}}=0$.
This dependence of the biexciton mixing on the Mn position is due to the different diagonal (Ising)
shifts of the bright and dark mixed biexcitons. Indeed, we find that if the Ising part of the exchange interaction is not taken
into account in the calculation of the interacting biexcitons the weights of the `exchange-free' biexcitons do not cross
(see the dashed lines in  Fig.\,\ref{B1}(a)). Therefore the dominant state does not change as $z_{{\rm Mn}}$ varies,
i.e. $\tilde {w}_{B-2;5/2}>\tilde {w}_{D-1;3/2}$ for any $z_{{\rm Mn}}$.

Figure\,\ref{B1}(b) shows the biexciton levels associated with the diagonal part of the Hamiltonian in Eq.\,(\ref{HTotal}),
whose dependence on $z_{{\rm Mn}}$ comes only from the Ising terms. These levels correspond to `exchange-free' biexcitons
so we denote them by ${\cal E}_{B-2;5/2}$ and ${\cal E}_{D-1;3/2}$. At $z_{{\rm Mn}}=5.5$\,nm the
gap $\delta={\cal E}_{D-1;3/2}-{\cal E}_{B-2;5/2}$ is essentially due to the hole-hole interaction and to
the magnetic anisotropy term $D_0{\hat M}_z^2$. We find that the intraband hole-hole Coulomb interaction induces a gap of
about 0.4\,meV between the bright and dark states. As the Mn atom approaches the
center of the cylinder the larger Ising shift pushes ${\cal E}_{B-2;5/2}$ above ${\cal E}_{D-1;3/2}$.

\begin{figure}[ht]
 \centering
 \includegraphics[width=0.49\textwidth]{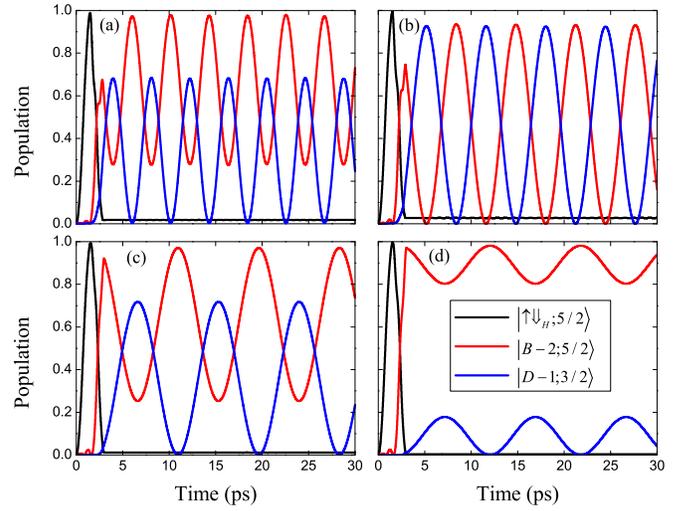}
 \caption{(Color online) Exchange-induced Rabi oscillations of bright and dark biexciton population
as a function of the Mn position. (a) $z_{{\rm Mn}}=0$, (b) $z_{{\rm Mn}}=2.1$\,nm, (c) $z_{{\rm Mn}}=3$\,nm, (d) $z_{{\rm Mn}}=4$\,nm.
$E=50$\,kV/cm.}
\label{OccupationB}
\end{figure}
The degeneracy point $\delta=0$ corresponds to $z_{{\rm Mn}}=2.1$\,nm and evidently coincides with maximal biexciton mixing in
Fig.\,\ref{B1}(a).

We now investigate the exchange-induced Rabi oscillations for mixed biexcitons. To this end we introduce the following setup:
(i) a sharp $\sigma_-$ pulse generates the heavy-hole exciton $|\uparrow\Downarrow_H;5/2\rangle$
(notice that the electron and the hole spin of this exciton cannot be flipped by the Mn-exchange interaction),
(ii) once this $\pi$ rotation is complete the pulse is switched off and one applies a second $\sigma_-$ pulse tuned to
the transition $|\uparrow\Downarrow_H;5/2\rangle\to|B-2;5/2\rangle$. Figure\,\ref{OccupationB} shows the Rabi oscillations of
the bright and dark biexcitons for several positions of the Mn atom. The oscillation amplitude reaches a maximum at
$z_{{\rm Mn}}=2.1$\,nm where the two biexcitons are strongly mixed (see Fig.\,\ref{B1}(a)). The
oscillations are much faster (the period ranges from 5 to 10\,ps) than the ones presented in the previous subsection, because the
Coulomb gap between the mixed biexcitons is much larger than the energy spacing between the LH excitons.

We stress that the effect of the h-Mn exchange alone can be singled out from the dynamics of the mixed biexcitons irrespective
of the Mn position on the $z$-axis, in contrast to the LH case where clear exchange-induced Rabi oscillations are observed
only if the Mn atom is closer to the cylinder edge.

The oscillations in Fig.\,\ref{OccupationB} were obtained without taking into account the intraband hole relaxation processes.
The separation between the LH and HH states (i.e the relaxation energy)
is around 7 meV which is below the frequency of the LO phonons in CdTe (this is typically around 22 meV - see e.g.
Ref.\,\onlinecite{Mizoguchi}. This means that in our case the relaxation is mostly due to LA phonons.
The reported values for this relaxation time range $\tau_h$ from 7-20 ps (see e.g. Ref.\,\onlinecite{Htoon}).

It is easy to see that two such processes are expected to damage the exchange-induced Rabi oscillations.
On one hand the HH exciton $|\uparrow\Downarrow_H;5/2\rangle$ can relax to $|\uparrow\Downarrow_L;5/2\rangle$; this effect can be reduced by a faster initialization of the HH exciton, that is by increasing the electric field. On the other hand,
 the dark biexciton $|D-1;3/2\rangle$ is depleted in favor of the purely LH bright biexciton
$|LL;3/2\rangle$. We find that this relaxation path is the main cause for
the damping of the Rabi oscillations.

\begin{figure}[ht]
 \centering
 \includegraphics[width=0.49\textwidth]{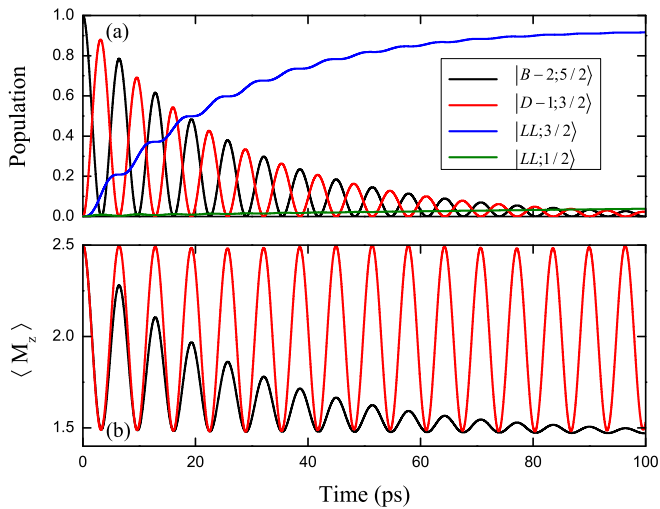}
 \caption{(Color online) (a) Damped exchange-induced Rabi oscillations of the biexciton populations in the presence
of the intraband hole-relaxations processes. The hole relaxation time $\tau_h=13.3$\,ps. (b) The average Mn spin as
a function of time for strong ($z_{{\rm Mn}}=2.1$\,nm) biexciton mixing. Red line - no hole relaxation, black line
- $\tau_h=13.3$\,ps, $E=75$\,kV/cm. }
 \label{Brelax}
\end{figure}

Damped Rabi oscillations of the mixed biexcitons in the presence of hole relaxation are presented in Fig.\,\ref{Brelax}(a),
where for simplicity we started the simulation with the fully occupied HH exciton state $|\uparrow\Downarrow_H;5/2\rangle$.
The hole relaxation time $\tau_h=13.3$ps.
The population of the $s$-shell biexciton $|LL;3/2\rangle$ increases smoothly on each oscillation of $P_{D-1;3/2}$
and is rather constant if the bright exciton is substantially populated. Eventually the system settles down to the
$s$-shell biexciton state ($t\sim 100$ps). We notice a small population of the state $|LL;1/2\rangle$;
this is due to the h-Mn interaction which couples the LH and HH spins through the transitions
$|D-1;3/2\rangle\leftrightarrows |LL;1/2\rangle$. The few meV gap between LH and HH exciton levels prevents a strong
coupling and explains the small and slow occupation of $|LL;1/2\rangle$. $P_{LL;1/2}$ is suppressed as the Mn atom approaches
the edge of the cylinder (not shown).

Let us stress that the rather fast switching between the mixed and $s$-shell biexciton requires {\it both} h-Mn
exchange and intraband relaxation. Indeed, in the absence of the Mn atom the transition
$|D-1\rangle\to|LL\rangle$ can only be achieved through hole spin-flip relaxation
which is a much slower process (i.e. nanoseconds \cite{Varghese}). Otherwise stated, the exchange interaction with a magnetic dopant
reduces the lifetime of mixed biexcitons.

The time-dependence of the Mn spin is shown in Fig.\,\ref{Brelax}(b). As expected $\langle \hat{M}_z\rangle$
displays periodic oscillations, the largest amplitude corresponding to the maximal biexciton mixing at
$z_{{\rm Mn}}=2.1$nm. Similar oscillations were obtained in a three-level model \cite{ReiterPRL2009}
in the context of all-optical manipulation of the Mn spin. However, the hole relaxation process strongly damages
the oscillations and the Mn spin is eventually frozen around 3/2.

\begin{figure}[ht]
 \centering
 \includegraphics[width=0.49\textwidth]{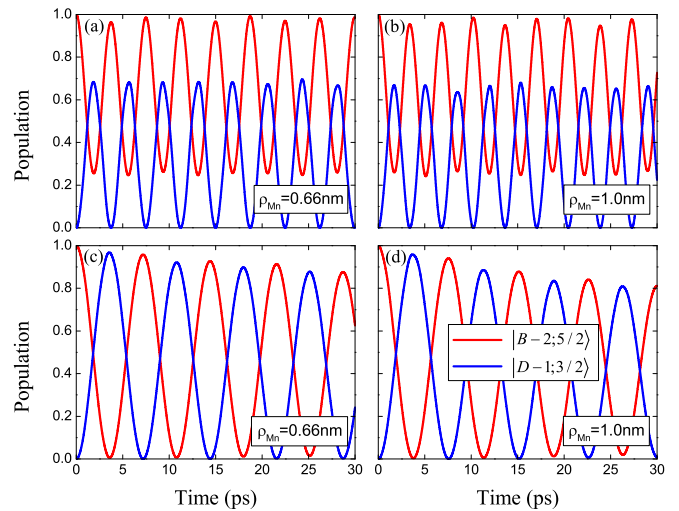}
 \caption{(Color online) The exchange-induced Rabi oscillations of the biexciton populations for different in-plane positions
of the Mn atom. (a) and (b) $z_{{\rm Mn}}=0$;  (c) and (d) $z_{{\rm Mn}}=2.45$nm. The initial state is $|B-2;5/2\rangle$ 
and the hole relaxation processes are neglected as in Fig.\, \ref{OccupationB}. }
 \label{inplane}
\end{figure}

Let us finally comment on the effects of the LH-HH mixing. In order to capture such 
effects we performed simulations for different in-plane Mn positions where the smaller weight $p$-shell 
components in Luttinger spinors in Eq.\,(\ref{spinors}) do not vanish. 
As long as  $J_{\pm 1}(\rho_{{\rm Mn}})\ne 0$ these 'minor' components also contribute to the exchange 
interaction and bring in more spin-flip processes. Consequently, the 'quadruple' structure of the LH excitons 
holds only approximately, in the sense that other smaller weight configurations contribute as well to a fully interacting state. 
We find however that the results on distinctive LH spin-flip processes as the Mn atom approaches 
the cylinder edges still hold. 


Figure\,\ref{inplane} shows that the exchange-induced Rabi oscillations of the biexciton populations
are still visible for non-vanishing radial coordinate $\rho_{{\rm Mn}}$ (i.e they are similar to the 
ones in Fig.\,\ref{Brelax}). Note that the populations
of the 'exchange-free' biexcitons $|B-2;5/2\rangle$ and $|D-2;3/2\rangle$ do not add to unity anymore as their 'minor'
components are exchange-coupled to other biexciton states with different $M_z$ as well (we did not include their populations 
in Fig.\,\ref{inplane}). The effect is negligible if $z_{{\rm Mn}}=0$ (see Figs.\,\ref{inplane}(a) and (b)) 
because the vertical confinement quantum number of the largest $p$-shell component of the 'mostly' LH Luttinger spinor 
is even (i.e $l=2$) and therefore the associated function $\xi_l(z)=\sqrt{\frac{2}{H}}\sin\left(\frac{\pi lz}{H}\right )$ 
vanishes. This does not happen at $z_{{\rm Mn}}=2.45$nm and a more pronounced decrease of the oscillation amplitude 
is noticed in Figs.\,\ref{inplane}(c) and (d). 
We obtained similar results for QDs of different sizes, provided the aspect ratio $H/2R$ leads to `mostly' LH 
ground state excitons.

\section{Conclusions}

We provide a theoretical study of the interplay of electron-Mn and hole-Mn exchange interactions in single Mn-doped
cylindrical QDs. The dependence of the exchange interaction on the manganese position is explicitly included in
the numerical simulations. Its effects can be traced from the multiple peak structure of the light-hole absorption
spectrum but also from the induced Rabi oscillations of the LH excitons and mixed biexcitons.

By changing the location of the manganese atom on the $z$-axis we identified two regimes of the light-hole exciton mixing.
The `quadruple' regime is characterized by simultaneous e-Mn and h-Mn exchange couplings, the interacting excitons being
typically made of four noninteracting states. This regime is suitable neither for the observation of the exchange-induced Rabi
oscillations nor for the preparation of an exciton with a dominant light-hole spin. The second regime is achieved
by placing the Mn atom closer to the edge of the cylinder, such that the e-Mn exchange is vanishingly small while the
h-Mn interaction is still important. In this case the fully interacting LH excitons are mostly made of two non-interacting
excitons which are mixed only through hole spin flip processes. Moreover, one recovers large amplitude Rabi oscillations
between bright and dark LH excitons.

We also investigated the dynamics of the mixed biexcitons with filled electronic $s$-shell and found hole-Mn exchange-induced
Rabi oscillations of the biexcitonic populations for any location of the Mn atom
on the $z$-axis. As expected, these oscillations are damped by the intraband spin-conserving hole relaxation.
The contribution of the Ising part of the exchange interaction on the biexciton mixing has been pointed out.
We hope that our results motivate further experimental and theoretical investigation of LH systems.

\begin{acknowledgments}
V.M, I.V.D. and R.D. acknowledge financial support from PNCDI2 program (grant PN-II-ID-PCE-2011-3-0091) and from grant
No.\ 45N/2009. B.T. acknowledges partial support from TUBITAK (112T619) and Turkish Academy of Sciences (TUBA).
\end{acknowledgments}


\end{document}